\documentstyle[11pt,epsf]{article}
\def\case#1/#2{\textstyle\frac{#1}{#2} }
\begin{document}
\title{\bf Lensing and caustic effects \\
on cosmological distances. }
\author{
{\sc G. F. R. Ellis$^{1}$,~ 
B. A.  Bassett$^{1,2}$,~
and~ P. K. S. Dunsby$^1$  }\\ \\
\normalsize{1~ {\it Department of Applied
Mathematics, University of Cape Town,}}\\
\normalsize{{\it Rondebosch
7700, Cape Town, South Africa.}} \\ \\
\normalsize{2~ {\it International School for 
Advanced Studies, SISSA - ISAS}}\\
\normalsize{{\it Via Beirut 2-4, 34014, Trieste, Italy.}} }


\maketitle
\begin{abstract}
We consider the changes which occur in cosmological 
distances due to the combined effects of some null geodesics passing through 
low-density regions while others pass through lensing-induced caustics.
This combination of effects increases observed areas corresponding to a given
solid angle even when averaged over large angular scales, through the additive
effect of increases on all scales, but particularly on micro-angular scales; 
however
angular sizes will not be significantly effected on large angular scales
(when caustics occur, area distances and angular-diameter distances 
no longer coincide). We compare our results with other works on 
lensing, which claim there is no such effect, and explain why the effect 
will indeed
occur in the (realistic) situation where caustics 
due to lensing are significant. Whether or not the effect is significant 
for number 
counts depends on the associated angular scales and on the distribution of 
inhomogeneities in the universe. It could also possibly affect the spectrum of
CBR anisotropies on small angular scales, indeed caustics can induce 
a non-Gaussian signature into the CMB at small scales and lead to  stronger 
mixing of anisotropies than occurs in weak lensing. 
\end{abstract}
\vspace*{0.2truecm}
\begin{center}
{\it Subject headings:\\ cosmology\,-\,gravitational lensing  \,-\,cosmic 
microwave background } \end{center}
\noindent

\newpage
\section{Introduction} \label{sec: Intro}
Cosmological angular\,-\,diameter distance and `observer area distance' 
(the latter equivalent up to redshift factors to the luminosity distance, see 
\cite{bi:elli71,bi:wein73}) lie at the heart of observational 
cosmology. They are used respectively to convert observed angles to 
length scales and observed solid angles to areas, at galactic distances and 
also on the surface of last scattering of the Cosmic Microwave 
Background radiation (CMB). 

They are equal to each other in the case where the universe is  
represented on large scales by a Friedmann-Lema\^{\i}tre (FL) universe with 
an exactly spatially homogeneous and isotropic Robertson-Walker 
(RW) geometry. This geometry is obtained as some kind of large-scale average 
of the 
manifestly inhomogeneous matter distribution and geometry on smaller scales 
\cite{bi:elli87}. The local inhomogeneity causes distortion of bundles of 
light rays 
and so alters the angular diameter distance and area distance through the
resultant gravitational lensing. Bertotti gave a power-series expansion
for this effect\cite{bi:Bertotti66}
while the Dyer-Roeder formula \cite{bi:dyer73,bi:schneider92} can be used 
at any redshift 
for those many rays that propagate in the lower density regions between
inhomogeneities.  However this formula is not  accurate for those ray bundles 
that pass very close to matter, where shearing becomes important. 

The case of weak lensing, where no caustics occur, has been studied in 
depth in the last few  years, including its effects on the CMB (e.g. 
\cite{bi:seljak}). Dyer  and Oattes (1988) \cite{bi:dyer88} included 
both shear and the varying 
Ricci term in a statistical study of lensing including caustics, and 
found that 
generic sources were demagnified, while a few sources were highly 
magnified. These results have been recently confirmed and extended by 
Holz \& Wald  \cite{bi:HW97}  and by Hadrovic \& Binney 
\cite{bi:HB97}. All of these  studies suggest that an 
average  source at high redshift in our universe will be demagnified due 
to caustics,  and hence that the area distance is not FL on average. 
That the area distance of the volume averaged inhomogeneous universe need not
be that of the underlying FL model is proven explicitly by
Mustapha, Bassett, Hellaby and Ellis \cite{bi:shrink2} and
discussed further by Linder \cite{bi:linder98}.

The usual assumption however, made explicit by Weinberg 
(1976) \cite{bi:Weinberg76} and accepted by most workers in the field,
see for example \cite{bi:Ehlers86,bi:seljak}
is that although the area distance will be inaccurately represented by 
the FL area distance formula on small angular scales due to the clumping 
of matter, when averaged over large enough angular scales that 
formula will be exactly correct, essentially due to photon conservation
\footnote{In \cite{bi:schneider92} (p. 133) areas are set to be equal as a fitting 
condition rather than supposedly arising from photon conservation.}.
It is our contention that this conclusion is wrong - areas 
will be different than in the corresponding FL universe both on small 
angular scales, and also when averaged to large angular scales. 
The increase occurs because if paths pass through underdensities (where the 
Dyer-Roeder approximation holds), they diverge more than in FL models; 
while if they pass close to matter, this will cause convergence which will 
often lead to caustics and  an associated divergence of geodesics, again 
resulting in  
an increase of area relative to FL models shortly after the formation of 
the caustic. This is essentially a consequence of  the non-commutativity 
of smoothing 
the geometry and calculating null  geodesics \footnote{See \cite{bi:ell84} 
for related 
discussion.}, or equivalently of fitting a FL background model and
determining geodesics. 

This paper explains the overall nature of the effect, giving geometric 
arguments as to why the combined effects will not average out to give the 
area distance associated with the underlying matter averaged FL model.  
We then 
explain why the  previous arguments either are  incorrect, or do  not 
apply to the real lumpy universe, once one follows 
light rays for long  enough that caustics have formed in our past light 
cone (which is a case of  considerable observational interest). 
We then give simple arguments as to how large the effect might be on different
angular scales.
While the effect associated with any
single lensing object is very small, there are a very large number of objects
in the sky that will cause lensing by the time our past light cone has
reached the surface of last scattering. 
The result of all the cumulative lensing on many scales is that the past light 
cone will have a fractal-like structure there. Thus caustics of
many scales will occur in all directions in the sky and the cumulative
effect on areas can be significant. 

The associated observational effects are complex, and depend on the 
model of matter distribution used and the  angular
scales observed.  On small angular scales, the distance covered on the last 
scattering surface for a given apparent angle in a lumpy universe will 
generically be more than in the corresponding FL universe model (which is normally 
assumed as giving 
the correct geometry), thus `shrinking' of images will occur 
- the apparent angular  size of a given object will be smaller than 
expected if lensing is not taken  into account, which will also affect
number counts on those scales. However due to the folding over of the light 
cone 
on itself associated with caustics, on larger angular scales the effect 
on angular sizes will average out - the observed angular sizes of large scale 
structures will be little affected, even though the associated 
areas can be quite different, because the light rays are little deflected when 
considered on  these scales; thus area distances and angular diameter 
distances will no longer be equivalent on these scales. This is 
consistent because the caustics cause the light cone to fold in on itself. 
Thus the resulting effect on particular 
observational relations will depend on whether it is overall angular size, 
or the associated observed areas, that is significant for the observations, 
as well as
on the angular sizes of averagings implied in the observations and 
resulting selection effects. Detailed calculation will be required to 
determine the magnitude of the effect in specific cases.

The effect is demonstrated explicitly by examples in paper II 
\cite{bi:shrink3}, and the principle at the heart of the effect 
is confirmed in an interesting rigorous way by analysis of exact 
axi-symmetric models in \cite{bi:shrink2}.
Together they show that photon conservation does not imply that  
the areas corresponding to a particular solid angle will be the same as those
in the background FL geometry, as  claimed in 
Weinberg's paper \cite{bi:Weinberg76}.

\section{Why lensing causes shrinking} \label{sec:shrink}
In general the relation between a scale $l$
perpendicular to the line of sight
at redshift $z$, and the angle $\theta$ it subtends when observed will 
depend on $\theta$, on the direction of observation (represented by a 
unit spacelike vector $\bf{n}$ orthogonal to the observer's 4-velocity), 
on the orientation of the arc, $l_{\perp}$, formed by the projection  of 
$l$ onto  the celestial sphere and implicitly on the angular scale 
$\theta_a$ over  which observations are  averaged,\footnote{In the case 
of the CMB, $\theta_a$ is the resolution of
the instrument. Detail smaller than this scale is lost.} as well as on 
the redshift $z$ of the object observed, viz:
\begin{equation}
l = r(z,{\bf n},\theta_a,l_{\perp}, \theta) \,\theta\;.
\label{eq:one}
\end{equation}
Here $r(z,{\bf n},\theta_a, l_{\perp},\theta)$ is the 
angular\,-\,diameter distance in 
a general universe, which will be anisotropic due to the shearing effects on 
the ray bundle, and for small angular scales will be independent of $\theta$. 

We expect that this anisotropy will tend to zero (as in the 
background FL model) as the averaging scale increases,  making 
the angular\,-\,diameter distance isotropic in the limit of large averaging 
angle, $\theta_a$ \cite{bi:linder98}. Thus a fairly good approximation for 
large\,-\,angle (e.g. {\it COBE}) experiments is that distortion 
(represented by large-scale shear in the null rays) is unimportant, 
and this is confirmed by weak lensing
studies \cite{bi:seljak}. However, this does not mean that the $r(z)$ 
converges to the FL distance corresponding to a given averaging of the 
geometry. Those paths passing through empty space between clustered
matter will be less focused than in the corresponding FL geometry
\cite{bi:dyer73}. On the other hand, sufficiently far down the null 
geodesics after passing strong lensing sources, conjugate points (and 
associated multiple images) will occur \cite{bi:seitz93}-\cite{bi:hasse}; the loci of
conjugate points in space time is a caustic sheet, a two-dimensional
surface to which the rays are tangent \cite{bi:nar}.
The typical behaviour of null rays near these caustics has been presented 
in \cite{bi:penr} (see Figure 49); the relation to gravitational lensing is 
discussed {\it inter alia} in \cite{bi:schneider92}. When averaged over a 
large angular scale, the combination of effects can lead to a change in 
the area-distance relation.  

Consider the past light cone $C^-(P)$ of the
space-time event `here and now', denoted by $P$. As a bundle of light rays 
${\cal B}(d\Omega)$ generating $C^-(P)$ (and subtending a solid angle $d 
\Omega$ at $P$) passes near a lensing mass $L$, the nearer rays are 
distorted in towards the central ray $\gamma_L$ linking $P$ 
to $L$. Radial ratios  will change (cf. \cite{bi:schneider92}, figure 2.3), 
decreasing as light  rays are bent inwards in the case of a spherically 
symmetric lens  (cf \cite{bi:gp}, Figure 2). The areas corresponding 
to a specific solid angle are invariant if the shear is small in a vacuum 
region, 
because transverse ratios will change in a compensating way,
but there will be a change in area if distortion is significant
or if there is matter present 
(as follows from the null Raychaudhuri equation, see e.g. [4,7])). 
Thus focussing is caused when strong lensing takes place,
 and this can be examined by 
ray tracing, by use of the geodesic deviation equation, or by
using the optical scalar equations. Consequently (see Figure 2 in 
\cite{bi:ref64}, or Figure 2.3 in \cite{bi:schneider92}), before cusps 
have formed, the area $dS$ of 
this nearby bundle of geodesics ${\cal B}(d\Omega)$ beyond $L$ will be 
less than if 
$L$ had not been there (i.e. in the reference background case, described 
by an exact FL geometry). Further out from the lens, where the density
is less than in the background, the effect will be reversed: areas 
will be larger. 

A crucial point here is that we must get the overall masses right. If we 
take a FL universe and add a mass concentration to represent some inhomogeneity
- a star, a galaxy, a galaxy cluster, or whatever - then the new universe
has greater mass than the old; so we expect the areas to be different simply
because the average mass density in a volume $V$ of the perturbed model 
that includes both $P$ and $L$, is different from that in the background model.
We need to correct the perturbed model to get back to the original mass in
this volume, so that the background model is correctly chosen to fit the 
perturbed model \cite{bi:elli87}. Or viewed differently, this is 
the requirement
that the perturbed universe can be obtained from the background universe by
rearranging masses while keeping overall mass conserved (this is the burden
of the Traschen integral constraints, \cite{bi:trasch}; when they are 
satisfied this is equivalent to correctly fitting the background model to 
the lumpy universe model, see \cite{bi:elljak}).
Thus when comparing lensing in a universe with given density $\rho_0 $
with that in the corresponding FL model, 
we must imbed the overdensity in an exactly 
compensating underdensity in order to maintain the value of $\rho_0$. 
The light rays in the outer underdense region will diverge more than 
in the background model, and those in the inner overdense region
will converge more. The standard view \cite{bi:Weinberg76,bi:schneider92}
is that these effects exactly cancel: the area in the perturbed model will 
be exactly the same as in the background FL model.

\begin{figure}
\epsfxsize = 3.8in
\epsffile{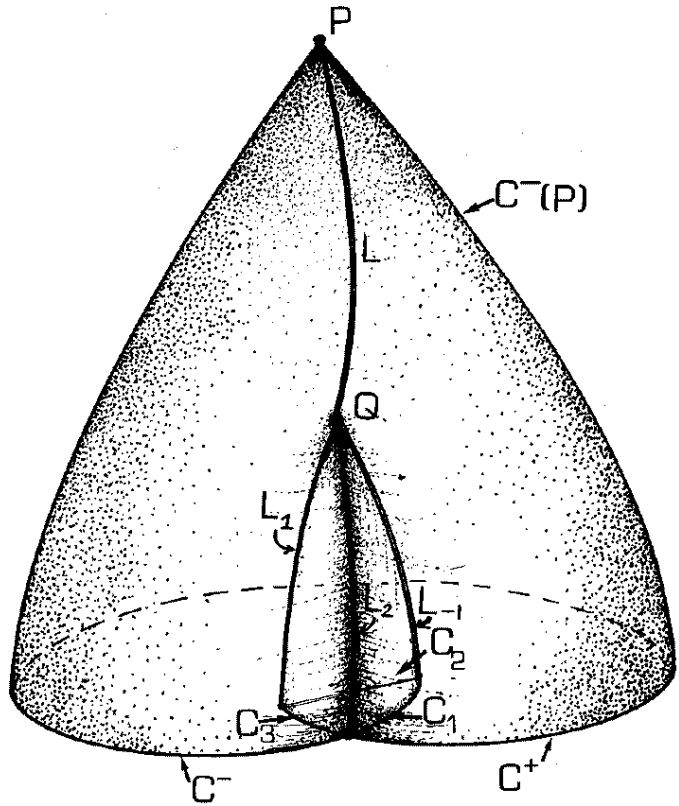} 
\caption{ A lens L and resulting caustics on the past light cone
$C^-(P)$ (2-dimensional section of the full light cone), showing in
particular the cross-over line $L_2$ and cusp lines
$L_{-1}$, $L_1$ meeting at the conjugate point $Q$. The intersection of
the past light cone with a surface of constant time defines exterior
segments $C^-$, $C^+$ of the light cone together with interior segments
$C_1$, $C_2$, $C_3$. 
}
\label{fig:onea}
\end{figure}

\begin{figure}
\epsfxsize = 3.8in
\epsffile{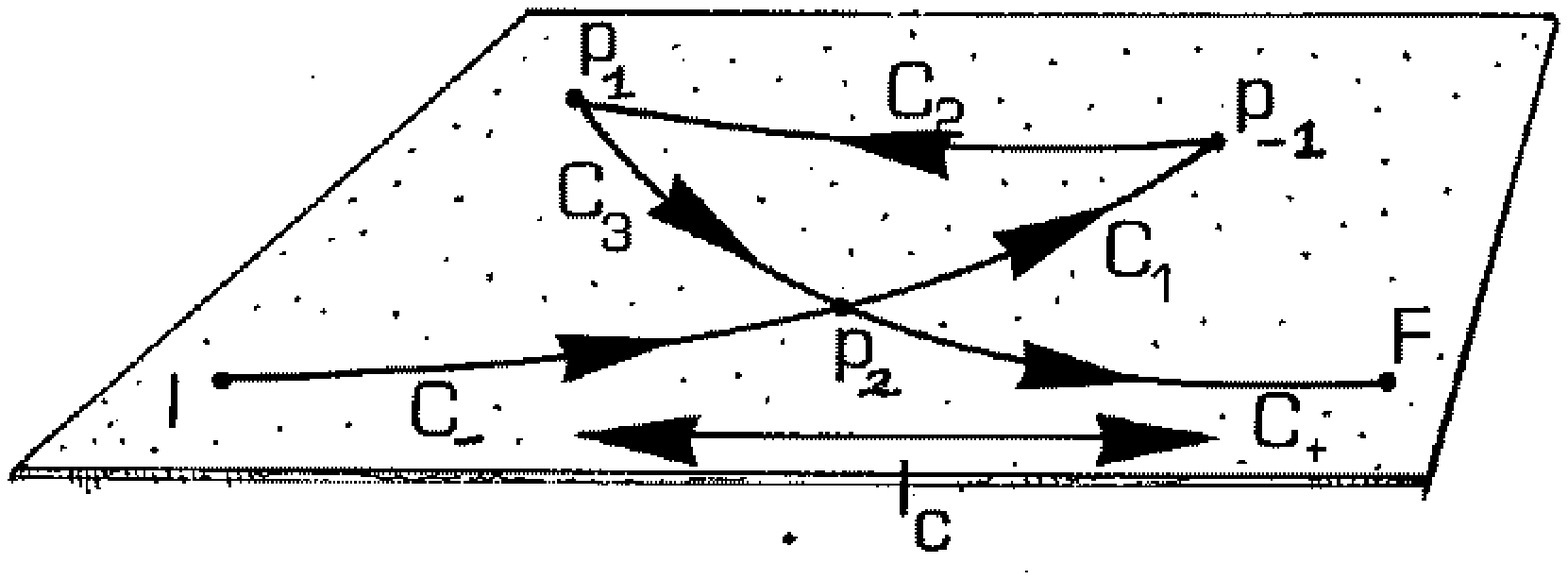} 
\caption{ The imaged
point moves forward along $C_1$ from $I$ to the cusp at $P_{-1}$,
backward along $C_2$ to the cusp at $P_1$, and then forward along $C_3$ to
$F$.}
\label{fig:oneb}
\end{figure}


However, this does not take caustics into account. 
After caustics have occurred, the null rays that were 
converging start diverging. Indeed at a caustic an 
infinite convergence is instantaneously converted to an infinite divergence
\cite{bi:seitz95}. Thereafter, both the rays that went through the 
less dense regions and those that went through more dense regions 
and were strongly lensed are 
diverging more rapidly than in the corresponding exactly smooth FL model. 
The only rays for which the area is less are those 
that passed close enough to a mass to be lensed so strongly as to
affect the area, but not close enough to form caustics and allow a
compensating re-expansion of the null rays to occur. As most rays are 
subject to greater divergence (see for example the simulations by 
Holz \& Wald  \cite{bi:HW97} discussed in Section 4), on average
the overall area (far enough down the light cone) will be greater 
than in the corresponding background model. Then on the 
corresponding angular scales, shrinking of images will occur.
Let $r_b(z,\Omega_0)$ be the
area distance of the background FL universe model with a value for $\Omega$
equal to that obtained by averaging the matter distribution appropriately.
Defining the pointwise shrinking factor $\gamma$ by $dl/d\theta$,
then for a finite angle $\Delta\Theta$ and corresponding distance $l$,
\begin{equation}
l = \langle \gamma \ \rangle r_{b} \,\Delta \theta\; = 
\langle \gamma \rangle l_{b}\,, 
\label{eq:onea}
\end{equation}
with average angular shrinking factor $\langle\gamma(z,\theta,\phi)
\rangle > 1$. 
Correspondingly there is a change in area: pointwise
\begin{equation}
dS = \beta\, r^2_{b} \,d\Omega = \beta dS_{b}\,, 
\label{eq:oneb}
\end{equation}
with average area shrinking factor $\langle\beta(z,\theta,\phi)\rangle > 1$ 
when averaged over some solid angle $\Delta\Omega$. 

As well known, there is no known covariant averaging procedure in General 
Relativity or agreed way of fitting a background model to the
real universe \cite{bi:elli87}, 
hence using different models for inhomogeneity and
associated averaging procedures will give different estimates for $r_b$
(this corresponds to the gauge freedom in fitting a background model to
the real universe, cf.\cite{bi:be89}). In paper II \cite{bi:shrink3} 
we study lensing  by local inhomogeneities which are 
explicitly chosen to satisfy matching conditions so that the total mass 
in a large sphere is the same as in the background model. 
In \cite{bi:shrink2} we use the standard astrophysical averaging - that 
on constant time slices in the synchronous gauge, ensuring that the mass
inside any inhomogeneous regions in this gauge
is the same as in the corresponding background model. 
An appropriate way of doing this, with explicitly stated assumptions 
on the potential $\Phi$, is set out in the paper by
Holz and Wald \cite{bi:HW97}. 

Given such a choice of fitting, we are interested in finding $\langle\gamma\rangle$, 
$\langle\beta\rangle$ in general, and in particular after caustics have occurred. 
Recent Hubble Space Telescope observations imply that
virtually everything beyond a redshift of 3 is at least weakly lensed, 
see e.g. the Hubble Deep Field
\footnote{website: http://www.ast.com.ac.uk/HST/hdf/} for 
between-cluster images 
\cite{bi:HDF1}, and many signatures of lensing are seen towards clusters, 
see e.g. \cite{bi:lens1}-\cite{bi:HDF2}.
At higher and higher redshift there will be more and more lensing. We 
are particularly interested in any effect 
this has on our past light cone by the time it has reached $\Sigma$, the 
surface of last scattering of the CMB, for this will influence our 
interpretation of the CMB data. The situation here is quite different than 
in relating lensing to discrete sources, for (in the instantaneous 
decoupling  approximation) the surface of last scattering is a spacelike 
surface;  thus we are interested in the relation of the real past light 
cone to a spacelike surface (in contrast to its relation 
to timelike lines, which is relevant in considering multiple lensing of 
discrete objects, see \cite{bi:schneider92}). Thus the issue is, 
What is the area $dS$ of a bundle of 
geodesics ${\cal B}(d\Omega)$ generating our past light cone $C^-(P)$ when it
intersects a spacelike surface $\Sigma$, or (almost equivalently), what is 
the distance $l$ traversed in this surface when one scans through 
an angle $\theta$? 
However there is an important subtlety here. 

\subsection{Distance traveled and distance gained}
The generic shape of a 2-dimensional section of the null cone occurring 
when simple gravitational lensing takes place, is shown in Figure 1 
\footnote{see also Figure 2 in \cite{bi:nit90},
Figure 5.1 in \cite{bi:schneider92}, Figure 4 in \cite{bi:fort94}, and 
Figure 25 in \cite{bi:ref94}}. 
Now consider, for fixed angle $\phi$, changing the direction of view ${\bf n}$ 
at $P$ through an arc ${\cal A}$ in the sky as the angle of observation 
$\theta$ increases continuously from some 
arbitrary initial direction $\theta_I$ to a final direction 
$\theta_F$, where the corresponding light rays pass through a 
transparent lens $L$ centred at $\theta_L$ ($\theta_I < \theta_L < \theta_F$)
\footnote{Thus this set or rays corresponds to moving radially relative 
to the lens image in the sky, rather than tangentially.}, and then develop 
caustics before intersecting the spacelike surface $\Sigma$. 
As the direction 
at $P$ continuously increases, the corresponding image point $p(\theta)$ in 
$\Sigma$ will move along the image of the arc ${\cal A}$ (a 1-dimensional 
curve) in the (2-dimensional) intersection of $C^-(P)$ with $\Sigma$, 
resulting in a series of forward, backward, and then forward motions 
because for each gravitational lens the 2-dimensional light cone section 
far enough 
down has at least two cusps and a cross-over (self-intersection) in it,
each of these being projections of the caustic sheet in the full-spacetime. 

Consider now the motion in $\Sigma$ of $p(\theta)$ as $\theta$ steadily 
increases from $\theta_I$ to $\theta_F$ (Figure 1b). Starting at the initial 
point $p(\theta_I) = I$ on $C_-$, it moves on $C_-$ from the left, 
through the cross-over point $p(\theta_{-2}) = P_{-2}$ (see Figure 2),
along $C_{1}$ to the cusp point $p(\theta_{-1}) = P_{-1}$, then back 
along $C_2$ 
through the lens point $p(\theta_L) = P_L$ to the cusp point $p(\theta_1) 
= P_1$, and then forward along $C_3$ through the cross-over point $P_2 =
p(\theta_2) = P_{-2}$ again and onwards on $C_+$ to the final point 
$p(\theta_F)= F$ 
on $C_+$. Hence it effectively traverses 
the same spatial distance (between $P_{-1}$ and $P_1$ along $C_2$) 
three times. 
If we choose $\theta_I$ and $\theta_F$ large enough, the points $I$ and $F$
will be essentially unchanged from where they would be in the background
model (these rays are essentially unchanged by the lensing mass). 

It will be useful to define two distances, both for the same 
angular change at the observer: we distinguish {\it distance traveled} 
$l_t$ along the full path: 
$$I \stackrel{C_-}{\longrightarrow} P_{-2}
\stackrel{C_{1}}{\longrightarrow} P_{-1} \stackrel{C_2}{\longrightarrow} P_1 
\stackrel{C_3}{\longrightarrow} P_2 \stackrel{C_+}{\longrightarrow} F \,,$$ 
calculated as a line integral along that path, and {\it distance gained} 
$l_g$ - how far the image point has moved in space from its starting 
point, calculated by determining the shortest distance between $I$ and $F$. 
This will be almost the same as the distance traveled along the path above
but omitting all the closed loop segments, i.e it is well approximated by 
the line integral: 
$$I \stackrel{C_-}{\longrightarrow} P_{-2} = P_2 \stackrel{C_+}
{\longrightarrow} F \,.$$ 
The difference is essentially that 
which occurs in a random walk - compare distance traveled by the agent 
(how far has his legs carried him) as against the distance moved (how far he 
is from where he started off). Both distances depend on the angle $\theta$, 
but the
first increases monotonically with $\theta$, while, for each angular scale
on which cusps occur, the second has a saw-tooth effect imposed on top of 
this uniformly increasing tendency. Because of this, the first
increases with $\theta$ on average much more than the second. For large
enough angles, the second will be almost the same as in the background
model (because the angular positions of $I$ and $F$ will be unchanged
by the lens); the backward travel due to cusps will almost exactly 
compensate for the extra forward travel they cause. Thus (in the case 
of a single lens) for large angular scales 
the distance gained will be almost the same as in the background; 
consequently (this distance being different from
distance traveled), this will not be true for the distance traveled.

\subsection{Addition of Areas}
What this shows is that after caustics have occurred, area distances and 
angular size distances are different. The former corresponds broadly to 
distance traveled, the latter 
to distance gained. A strongly-lensing object $L$ will cause 
caustic lines on $\Sigma$, defined as the intersection of the caustic sheet 
with $\Sigma$. These will be spherically symmetric if the lensing object is 
spherically symmetric, and will be centered on the null geodesic $\gamma_L$ 
from $P$ through $L$ to $\Sigma$; similarly the critical curves (the images
in the lens plane of the caustic lines) will also
be circles around $\gamma_L$. Considering 
the full two-dimensional intersection ${\cal S}$ of $C^-(P)$ with $\Sigma$, 
in the spherically symmetric lens case,
it will be given by rotating the 1-dimensional picture (Figure 1)
about the central geodesic $\gamma_L$. The result is an eggcup-shaped 
section of the past light cone moving in to the interior and centred 
on $\gamma_L$. Defining the cusp angle $\theta_c$ by
\begin{equation}
\theta_c = \theta_1 - \theta_L
\end{equation}
(see Figure 2b), 
the area $A_c = \pi r_b^2 \theta_c^2$ is covered 3 times by any solid angle
centred on $\gamma_L$ of angle greater than about $3\theta_c$, where we use 
the background area distance $r_b(z)$ to convert angles to distances\footnote{
Actually we should rather use a modified distance estimate that takes 
distortion and consequent changes in area distances due to lensing 
into account; here we ignore that extra complication, but
it will have a significant effect if strong lensing takes place.}. 
Thus the real area corresponding to an angle $3\theta_c$ centred on 
$\gamma_L$ is about $3A_c$ whereas in the 
background model it will be simply $A_c$; so the area shrinking factor 
$\langle\beta\rangle$
will be about $3$ for such angles. However for the corresponding angular
size factor $\langle\gamma\rangle$ we will find 
$\langle\gamma\rangle \simeq 1$ because the lens will 
already have little effect at angular separation
 $3 \theta_c$ from $\gamma_L$. 

To work out the area relations properly, we need to use the determinant 
$J$ relating solid angles at the observer to areas in the source plane 
\cite{bi:schneider92}. The key point here is the sign of $J$: the regions
where angular travel is forward as discussed above will correspond to
regions where $|J| > 0$; the regions where angular travel is backwards
correspond to where $|J| < 0$. Thus in adding up areas, we have two
options: adding up the magnitudes of areas (where we assign a +ve value to all 
areas, i.e. we integrate $|J|$ over the relevant solid angle)
or adding up signed areas (where we assign a -ve value to areas
where $|J| < 0$, i.e. we integrate $J$ itself over the relevant solid angle). 
The former corresponds 
to distance gained, the latter to distance traveled. 

It is the latter
that is relevant to number counts, for they depend on the total
area occurring irrespective of the sign of $J$, and it is this we use to
define area distance in the realistic universe model, and hence to 
determine the area ratio $\langle\beta\rangle$. 
Hence in equation (\ref{eq:oneb}),
we assume all signs are positive (i.e. we take the modulus of areas and 
solid angles in calculating $\langle\beta\rangle$). The claim is that when the background 
model is properly matched to a more realistic lumpy universe model, we 
will find $\langle\beta\rangle >1$ on averaging over large angular scales.

\section{Response to Weinberg's arguments}
The paper by Weinberg \cite{bi:Weinberg76} explicitly considers this averaging 
issue, and argues that there is no overall such shrinking effect. He gives two 
independent arguments as to why this is so; clearly it is necessary
that we answer them here. 

The first point is that Weinberg's paper
does not explicitly take into account the effects of caustics, which we
are identifying as important.
His first argument is by explicit calculation (based on the 
previous work of Gunn and Press \cite{bi:gp}) of bending by a single
finite-radius clump of matter, and of the resulting intensities.
However he only allows for two ray
paths from the source to the observer - whereas in
the generic case there will be three such paths. To first order in $q_0$ 
(i.e. assuming $\Omega_0 \ll 1$) he 
finds that the luminosity distance (estimated from the combined intensities
of the two images) is the same as in the FL model. If we include
the general third image we may expect a different result. Additionally the
estimates used are only valid for $z \simeq 1$ (\cite{bi:gp}, p.400), and 
hence do not cover the large-z case we are interested in.

He then gives a second argument, based on photon conservation. This argument 
is correct in that it determines the average number of photons intercepted 
by a telescope in terms of the area of a sphere drawn about the object, and
works on the basis that this number is conserved (a good approximation in the
context considered). The problem is that Weinberg then {\it assumes that the 
area of this sphere can be
calculated from the FL area formula}, whereas this is precisely the issue
in question. At first glance one might think the answer is obvious
because here we are dealing with the up-going future light cone from the
source, rather than the down-going past light cone from us, and at late times
the universe is very similar to a RW universe; but by the
reciprocity theorem, these light-cones are essentially equivalent to each 
other. Just as the past light cone of the event `here and 
now' will develop numerous caustics as we go further into the past, so 
will the future light cone of the 
source as we go further to the future from that source
provided it is far away enough in the past (if this
were not so, multiple images of the same source could not occur); and 
the sources we are concerned with, when dealing  with the CMB, are very far
away - on the surface of last scattering.
Just as our past light cone develops a hierarchically structured
set of caustics by the time it reaches a source $S$ on the surface of last 
scattering, so the future light cone of the source $S$ will have developed 
a complementary hierarchically structured set of caustics by the time it 
reaches us. The area of this future light cone at the present time therefore
cannot be assumed to have the FL value; indeed this is essentially the 
quantity we have to calculate. Thus the argument in Weinberg's paper does
not establish the result that the averaged area distance will be the same
as in a FL universe, as claimed; it effectively assumes this result, by 
assuming this area is equal to that in a FL model.

Indeed on reflection it becomes clear that while photon  conservation 
leads (via the reciprocity theorem) to the important result that  
lensing does not affect radiation intensity, it 
cannot determine the  cross-section area of the past null cone and hence 
area distances, for that is determined by the Einstein field equations
(specifically, by the null Raychaudhuri equation).
Given that this argument does not work in the case of strong lensing, 
when caustics occur, it is clear that it does not work in the case of 
weak lensing either. In both cases photon conservation relates 
measured intensities to the area of the past light cone, but cannot 
determine the latter, which is determined by the matter present via 
the gravitational field equations.

\section{The real past light cone}
In the real past light cone, many light rays - even if passing through
galaxies - will pass through low density
regions all the way back to the surface of last scattering 
and so will have a larger area than
in a FL model; the Dyer-Roeder formula will apply to them. Many others
will pass near matter clumped on different scales and may be 
strongly lensed; this will then result in an area increase due 
to the occurrence of caustics, as outlined above.
The additional area (about $2A_c$ for a spherical lens) will be very small
for any particular lens, because cusp angles are small (between $3''$ and 
$30''$ for realistic astrophysical objects). But the point 
is that the number of lensing objects is very large. Each star will 
cause lensing, acting as an opaque lens\footnote{and substituting
its own radiation for the background radiation within the angular size
of its opaque disc, \cite{bi:stef95}.}, as will massive planets; each 
sufficiently concentrated galaxy core will cause lensing, acting as a 
transparent lens, as will each sufficiently dense cluster of galaxies
\footnote{Voids with sufficiently sharp edges can also cause lensing, for 
they are equivalent to using the usual lensing equations with an effective 
negative mass density; however probably actual voids will not have 
sharp enough edges for this to occur.}. 

\begin{figure}
\epsfxsize = 3.8in
\epsffile{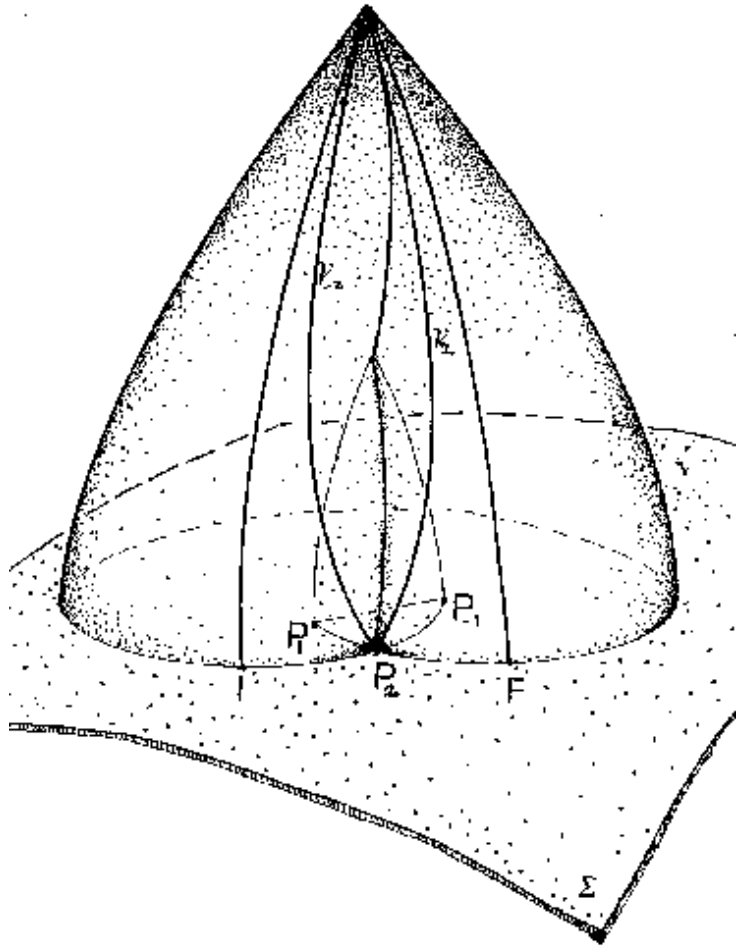} 
\caption{ The unique geodesics defined by a spacelike surface $\Sigma$ 
intersecting the caustic structure. Geodesics $\gamma_{-2}$, 
$\gamma_2$ joining $P$ to the cross-over point $P_2$ and defining the
cross-over
angle $\theta_2$. }
\label{fig:twoa}
\end{figure}

\begin{figure}
\epsfxsize = 3.8in
\epsffile{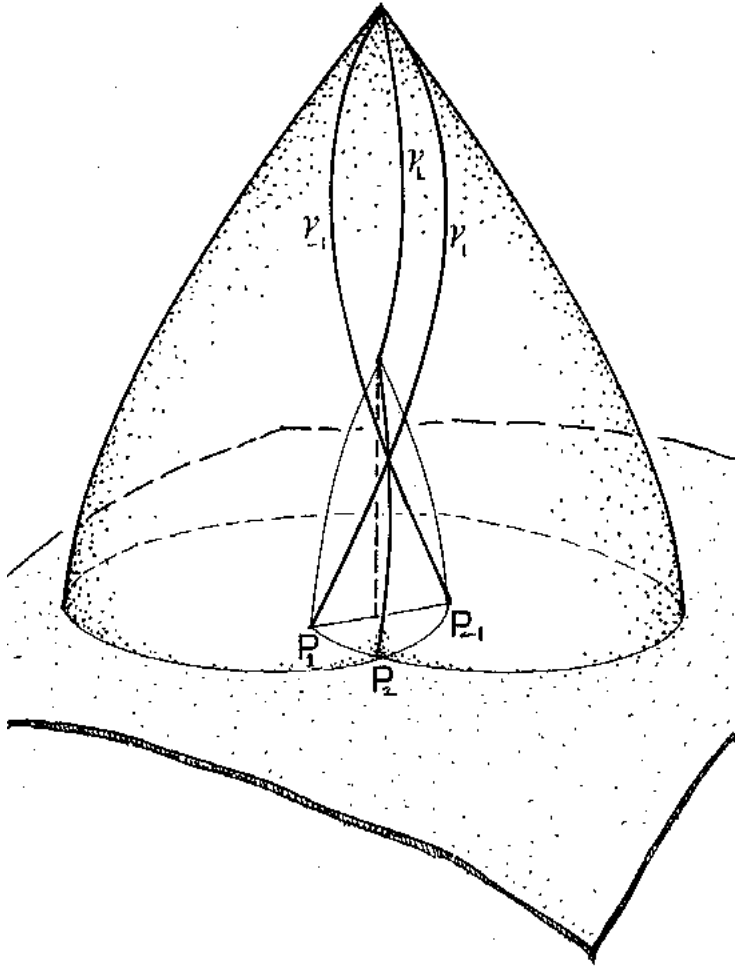} 
\caption{Geodesics $\gamma_{-1}$, 
$\gamma_1$ joining $P$ to the caustic points $P_{-1}$, $P_1$ and
defining the cusp angle $\theta_1$. Also shown is the central geodesic 
$\gamma_L$. }
\label{fig:twob}
\end{figure}


In many cases the lensing will cause caustics to 
form, indeed often this will happen quite close to the lensing mass; 
for example 
in the case of the sun, bending of light by $1.75''$ at the limb will cause a 
caustic to occur
in initially parallel light rays at that distance where 
the sun subtends an apparent size of $3.5''$ - which is $.0093$ parsec 
or $.03$ light years, and so much less than inter-stellar distances. 
Once a caustic occurs in our past light cone, further lensing (caused by 
inhomogeneities further down the past light cone) can never remove it, 
but can introduce new caustics. Furthermore, a single object may cause 
multiple caustic sheets; for example, sufficiently far down the past light 
cone, an elliptic lens will cause the double-caustic pattern noted by various 
workers \cite{bi:bland86,bi:nar}. 
      
Hence the number of caustics in our past light cone, by any high redshift
and in particular by the time it reaches 
the surface of last scattering, will be extremely large, of the order of at 
least $10^{22}$, and will occur in a hierarchically structured way
with larger cusps (due to galaxies and clusters) superimposed on smaller 
cusps (due to stars and planets), leading to something like a fractal 
structure. It is important to realize that as we are interested 
here in effects on very distant number counts or on the CMB spatial spectrum,
rather than in detailed lensing positions related to specific sources, 
{\it there is no alignment problem}: the surface of last scattering $\Sigma$ 
effectively occupies  the entire sky; and most detectable objects will 
cause caustics by then at least on small angular scales, 
because $\Sigma$ is a very large distance away, 
corresponding  to a redshift of about 1200 and most of these objects are
made up of density concentrations like stars that will cause strong lensing. 
In addition any particular inhomogeneity may contribute to multiple cusps 
on different scales: multiple counting of the effects of any particular
mass element is appropriate when a star causes micro-cusps and is situated in
a galactic core which causes larger cusps, in a galaxy in a dense cluster
which in turn causes even larger cusps. The individual stars then contribute 
to the formation of cusps on all these scales. 
Thus it is likely that an appreciable fraction 
of the intersection of our past light cone with $\Sigma$ will be covered by 
at least a single caustic.

Considering this fractured structure of the real past light cone 
$C^-(P)$ by the time it hits the surface of last scattering, it is clear 
there are potentially significant effects on the overall area 
resulting from the cumulative effects of all lenses. The overall effect will 
remain even after the averaging over a large angular scale due to convolution 
of the incoming information 
with a detector point spread function, because (unlike the angular distance)
addition of areas is additive; the integrated magnitudes of area increments 
will continue accumulating as we consider larger and larger scales, although
signed area increments will 
approximately cancel out if the model is approximately RW in the 
large\footnote{This is in effect a legitimate version of the argument 
put forward by Weinberg.}. We argue that 
distance traveled (or area distance) is substantially affected when
averaging on any angular scale, but that distance gained (or angular 
diameter distance) is significantly affected up to some angular scale 
$\hat{\theta}_c$, but not much affected on larger angular scales. 
The value of $\hat{\theta}_c$ depends 
on the clustering of 
matter at all redshifts up to the surface of last scattering; for a 
single spherical lens it is about $3 \theta_c$. 

\subsection{Simple estimates}\label{sec:simple}
To estimate the relation between the various distances on the
 surface of last scattering $\Sigma$, we note that the rays passing through
empty space or through uniformly distributed matter far enough away
from inhomogeneities
will correspond to the Dyer-Roeder distances. Thus the first issue here
is {\it what fraction of the sky will correspond to rays that have
passed only through empty space away from clustered matter, as a
function of redshift?} 
The problem here 
is that there is a hierarchically structured answer to this question: the
response will differ dramatically depending on the angular scale involved. 
For example on a microscale most light rays passing through a star cluster 
or galaxy pass through empty space (the cross section for collision with
a star being something like $10^{-5}$ or less), whereas on the galaxy
scale these rays are passing through smoothly distributed matter. Thus
this fraction may be very high at small angular scales but almost zero
at large angular scales.

The second issue is the effect
of strong lensing. First consider
the situation of a single lensing object producing a pair of
cusps in the radial intersection of the light-cone with $\Sigma$. 
The key issue here is what is the angular size of the cusp 
separation at last scattering, 
i.e. what is the angle $\theta_s = \theta_1-\theta_{-1}$ between the two rays 
that reach the outer edges $P_1$, $P_{-1}$ of the caustic at $\Sigma$ 
(Figure 2a). Closely related is the angular separation 
$\theta_m = \theta_2 - \theta_{-2}$ of the two rays 
that intersect $\Sigma$ in the self-intersection where the light
cone folds in on itself (Figure 2b). 
This will be the angular separation of multiple images 
of a single space-time event in the surface $\Sigma$;
its value will be approximately $2\theta_s$. 
Point sources, represented by timelike worldlines, move with the
fundamental 4-velocity, and can be multiply imaged up to an angle 
$\theta_M \simeq (3/2) \theta_s$ from the centre. This maximum lensing angle
can be of the order of $10''$ for nearby galaxies and $30''$ for 
galaxy clusters, for we have already seen deflections or arcs on these
scales, but could be larger, for two lensed images in the rich cluster 
AC 114 are separated by $50''.6$
(see \cite{bi:lens3}) and multiple lensing may increase the effective
angle significantly (see below). However many galaxies
and clusters will - as coherent objects -lie below the critical surface
density needed to create cusps; the stars of which they are made will 
nevertheless cause smaller scale cusps.

Consider then a distribution of such objects, but still only taking into
account single lensing (light rays only pass close enough to one such
object to be appreciably deviated). Then when we consider some angular scale
$A \gg \theta_s$, images on those scales will be negligibly affected
by the lensing. The effect is like wrinkled glass: small scale structure is 
blurred but large scale structure behind is reasonably clearly 
visible. We can immediately  attain a simple
estimate of the relation between the various distances mentioned above: 
the background distance $l_b$ will be well approximated by $l_g$ 
for such scales, with error
at most the distance $l_c$ corresponding to the angular scale $\theta_c$, 
because the distances between the widely separated rays will not be affected
by more than this amount. However $l_t$
will be different: from the argument above, for each spherical lens it 
will be increased by approximately 2 times the distance corresponding
to $\theta_c$, because that path will be traversed 3 times as $\theta$ 
increase from $0$ to $A > 3 \theta_c$ \footnote{If strong lensing takes 
place, the distance can be much greater, because then the (local) area 
behaviour in the lens will be quite different than in the background model.}; 
the corresponding area will be triply covered, so the area will be increased 
by twice the area of a disk of angle $\theta_c$. This will occur on top of 
an increase of area resulting from the light rays at the lens having passed 
through empty space up to the time they reached the lens. For the double 
caustics of elliptic galaxies, there will be an increase by a factor 3 
between the inner and outer caustic, and a factor 
5 within the inner caustic lines. 

This will be true for each single caustic line
encountering the surface $\Sigma$. Thus the issue is {\it what fraction 
of $\Sigma$ will be covered by single or multiple caustics?} Equivalently, 
the effective shrinking factor will correspond to the degree of multiple 
covering of $\Sigma$ \footnote{The degree of multiplicity equals the number 
of images of the source in the ``unperturbed direction".} 
by caustic surfaces \cite{bi:dyer88} which in turn 
equals the number of sources in the unperturbed direction. Defining the 
multiplicity of covering $M$ as the number of times the same segment of 
$\Sigma$ is traversed due to multiple caustics: 3 for a simple caustic
as in a spherical lens and the outer region of elliptical lenses, 5 for 
the interior of elliptical lenses, and so on, what we are interested in is:\\ 

(a) how does $M(\theta,\phi)$ vary over a surface $\Sigma$ of 
constant redshift?

(b) what is its average value $\langle M \rangle$ over $\Sigma$?

(c) How does $\langle M \rangle$ vary with redshift $z$?\\

The value of $\langle M \rangle$ can be very high behind dense
clusters of galaxies, as many caustics will overlap there; the occurrence of
arcs at angular scales up to about $10'$ confirms the multiple imaging
occurring in these cases on those scales. However these clusters do not 
cover a large fraction of the sky; so the issue is what is its value in those
areas of the sky between such galaxy clusters? What fraction of rays
pass through low density areas where convergence is less than in FL models?
 
The increase in area due to the combination of low-density light propagation
plus caustics can be significant, as is  
supported by recent numerical studies of 
strong lensing. In the extreme limit of point-mass objects, Holz \& Wald 
\cite{bi:HW97} give results showing that\footnote{When lensing occurs, 
focussing and then re-expansion results in a loss of area relative to the
background model from the start of focussing until the re-expanding light 
rays have regained the lost 
area.  From Holz and Wald one can see for example that in the $z = 3$ case, 
17\% of the beams have a magnification with amplitude less than one; these
are the beams where one is losing out overall. On the positive area side, in
terms of the percentages across the horizontal axis in Figure 5 of HW, 
the loss from 35\% to 47\% is recouped by 63\% (i.e. the area under the curve
from 35\% to 63\% is the same as in a FL model); from 63\% to 100\%
is all gain with an average gain factor of 2.  The total are under the curve
on the positive area side (i.e. for greater than $35\%$) is $102$, whereas the
corresponding FL area is $65$; the average amplification factor is thus
$102/65 = 1.4$. The average factor will be the same on the negative side 
because of the overall balancing of signed areas.}
 for the average  over all of 
the photon beams, 
$\langle\beta\rangle
 = 1.1$ at $z = 0.5$ and $\langle\beta\rangle
 =  1.2$ at $z =  2$ in an $\Omega = 
1, \Lambda = 0$ universe while $\langle\beta\rangle
 =  1.4$ at $z =  3$ in the same 
model. Here $\langle\beta\rangle$ is  the increase in area  of the
wavefront over that 
in the background FL model, see Eq.  (\ref{eq:oneb}). This shows that the 
increase in area  can be large  due to the combination of rays traversing
low-density regions and the existence of caustics, and is a 
rapidly  increasing function  of redshift. Their study further showed 
that  while only $6\%$ of beams  had developed caustics by $z = 0.5$, 
$26\%$ and  $37\%$ of beams had  developed caustics respectively by 
redshifts $z = 2$  and $z = 3$ in the $\Omega = 1, \Lambda = 0$ cosmology. 

These can be considered as upper limits for lensing at one scale,
since they correspond to the case of point masses. 
The important issue then is that galaxies are made of
point masses - stars - on top of the smoother dark matter halo, 
which itself contains a compact MACHO distribution. 
Thus the Holz and Wald results represent reasonable estimates for the area
amplification factor due to microlensing in galaxies. What fraction of the
sky is covered by galaxies? 
In a recent study, Premadi, Martel \& Matzner \cite{bi:PMM97}  modeled 
the large  scale matter distribution using a realistic P$^3$M N-body code 
with extended resolution for locating galaxies, and found that all  photon 
beams intersected a galaxy by $z = 5$ if $\Omega = 1, \Lambda = 
0$. In open or $\Lambda$-dominated cosmologies the corresponding redshift 
was less, of order $z \sim 3$. Intersection with each galaxy ensures caustics 
due to microlensing  by stars adding significant area to the wavefront: 
between 10\% in a low density universe, or 40\% in a high density model. This 
is not removed by angular averaging, i.e. it is not important that 
telescopes cannot resolve individual  microlensing effects for the 
average area distance to altered. 

A number of effects alter these basic estimates. First Holz and Wald 
do not take their estimates out
to nearly the redshift we have in mind (up to say $z = 1200$). The area factor 
could increase greatly in this distance - say up to about 3. 
Secondly this only treats the micro-lensing contribution to cusps but
galaxies themselves and clusters will also contribute in many cases
(on larger angular scales) due to the core and/or halo densities. 
This effect may be somewhere from 5\% to
30\% increase in area. Consider for example the contribution from the 
cores of an isotropic population of blue galaxies with $200,000$ images per 
$deg^2$ per $mag$, giving $30$ per $arc\, min ^2$ at $29$ mag with 
redshifts in the range 1 - 3. (Tyson {\em et al} \cite{bi:tyson}) 
A small fraction of these objects at redshifts below 2 form caustics. However, 
the cores of star forming regions may have masses of the order of $(1-2)\times 
10^{10} M_{\odot}$ or more within a radius of $2kpc$ \cite{bi:pettini}, and may 
substantially alter areas of light bundles that pass close to them. Galactic 
cores at redshift 3 or higher subtend angular diameters of $(2\, -\, 3)''$, and 
if conditions are conducive for multiple imaging, the total (core + caustic) 
angle is about $(10\, - \, 12)''$ (the core + cusp angle is half the caustic angle), 
so that the each lens has a cross-section of about $17\, -\, 18$ square 
arcseconds. For the population of blue galaxies at $29$ mag alone, this totals 
to $4.6\%$ of the sky  covered with caustics due to lenses between $z= 1$ 
and $z=3$,  so that the covering factor amongst the population of blue 
galaxies  is  $109.2\%$ for $\Omega = 1$. At a redshift of $5$ the 
covering is about $115\%$; and we are interested in what happens by the 
time we reach the surface of last scattering. 

The way we model the matter distribution is crucial.
Ignoring the point-like masses in galaxies and 
treating only the smooth component, the effects of caustics become almost 
negligible, even at $z \sim 3$ \cite{bi:HW97}. In some sense this is 
obvious though, since spatial averaging in the limit must remove all 
lensing effects. However, this averaging is unphysical. In the real 
universe microlensing will take place in each galaxy and 
increase the actual area of the past light cone significantly, and on 
top of this we must allow for any increase
due caustics caused by galaxy cores and galaxy clusters. 

There is another, extremely model-dependent complication: that of the 
effects of {\it multiple small-angle scattering} between $P$ and $\Sigma$. 
When this takes place, there 
are two effects: firstly, this can introduce new caustics in the
past light cone structure (but cannot remove any that already exist).
Secondly, it will alter the angular size of existing caustics,
leading to a random walk in the effective angle $\theta_c$ for a given lens,
potentially leading to overall deflection distances that can be quite large 
if sufficient such scatterings take place. 
How large depends on the number of scatterings and angle of each one, in turn
depending on the distribution of inhomogeneities all the way back to $\Sigma$,
but they can potentially correspond to angles of $\sim 1^\circ$ 
\cite{bi:fuk}.

It is clear then that in a realistic model of the universe, the past light
cone is an extremely complex object covered with cusps on many angular scales.
The probability distribution for $\theta_c$ will be peaked  at angles from 
microarcseconds to at least 
$30''$, but may extend up to $30'$ or so because 
of multiple scatterings combined with the effects of superclusters, which 
could be significant \cite{bi:bark96}. There 
will be a tail up to larger angular scales due to black holes, but of 
very low amplitude. We estimate an area shrinking factor, when 
averaged on large  scales (or over the whole sky), of between 1.1 and 3, 
at the surface of last scattering; it could be greater. 

\section{Observational effects}
The effect on observations could be appreciable at some angular scales
once the cumulative effect of lensing has started to build up - at $z >  3$
and beyond. In measurements that depend on area effects, the increase in 
area due to shrinking will broadly correspond to the multiplicity $M$. 
The actual observational effects will depend on the 2-dimensional 
distribution of cusps on surfaces of constant redshift such as $\Sigma$, 
which cannot easily be estimated from  the 1-dimensional projections 
considered here. The figures obtained from such studies should correspond to 
those obtained by considering the rays propagating through low density 
regions only, because of the overall necessity to average out to a
FL geometry on large scales.

Number counts will be altered when $\langle\beta\rangle > 1$ because the areas 
covered by the light rays in a given solid angle are larger than estimated 
from the FL formula, by the area shrinking factor 
$\langle \beta \rangle ~\simeq ~\langle M
\rangle$; however the detection probability will be lowered and this
will tend to compensate. This is taken into account already
in detailed lensing studies, but not perhaps in all high-z number count
analyses where it might make a difference at the few-percent level.

The angular correlations of CMB fluctuations will also be affected by
strong lensing, and is expected to cause much stronger alterations 
than occurs with weak lensing. However the effect is not just an 
alteration of apparent scale, because the distance traveled along the 
surface  of last scattering $\Sigma$ by the 
measuring beam is the distance traversed $l_t$ (cf above), which is 
greater than the distance gained $l_g$  (the extent of each pair
of cusps is  traversed three times, rather than once). 

\begin{figure}
\epsfxsize = 3.8in
\epsffile{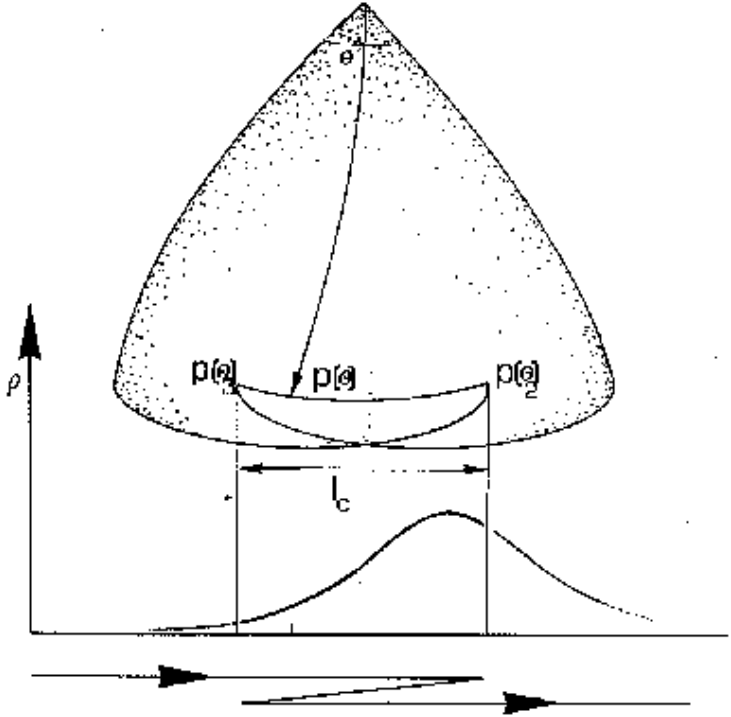} 
\caption{ 
Sampling of an inhomogeneity by the 
observational point 
at a caustic results in a change of the observed profile, because the
same region of the profile is traversed 3 times by the observational point. 
}
\label{fig:threea}
\end{figure}

\begin{figure}
\epsfxsize = 3.8in
\epsffile{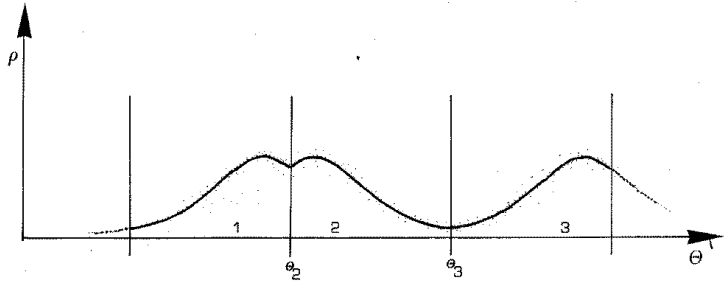} 
\caption{ A Gaussian profile will not remain a Gaussian profile.}
\label{fig:threeb}
\end{figure}


If we consider an observer sweeping a narrow beam across the sky and
measuring incoming radiation in that direction, at the surface of last 
scattering this beam will traverse the cusps that occur in the intersection 
of the past light cone with the surface of last scattering, consequently 
moving forward, backward, and then forward each time such a cusp occurs
[see Section 2 above] and almost performing a random walk when one takes 
into account the whole hierarchical structure of these cusps. 
Thus any particular small-scale temperature fluctuation will be sampled 
several times as it is scanned both forwards and backwards by the measuring 
beam; hence any Gaussian fluctuations on these scales will be 
measured as  non-Gaussian on these scales; in effect, the actual spatial 
distribution is convolved with the saw-tooth sampling pattern.
This will induce non-Gaussianities in the CMB anisotropies at the scales 
of the largest caustics [Figure 3]. 

What is measured on large 
scales is determined by the distance gained $l_g$, which tells us 
when the sampling point reaches new large-scale features of the 
inhomogeneous distribution of matter on the last scattering surface. The 
smaller backward and forward traverses are then averaged over in an 
effective coarse-graining. The corresponding shrinking factor relative to 
the background will be significantly different from unity on scales smaller 
than the peak in the distribution of $\theta_c$ over the sky,
but will be close to unity on scales rather larger this scale. 
To determine this distribution requires detailed modeling. 

\section{Conclusions } \label{sec: concl} 
In this paper we have asked the question of how strong lensing with 
multiple caustics due to  inhomogeneity  will change cosmological distance 
estimators and the total area of the past null cone. 
This is much more complex than the case of weak lensing since it involves both 
nonlinearities in the matter distribution {\em and} non-perturbative, singular 
developments in the past null cone (even if the light-bending is weak
in the sense that only small-angle scattering occurs).   

We find that the angular-diameter distance is not strongly affected on large 
angles but that the {\em all-sky averaged} area distance is significantly 
increased 
due to the folding of the null cone after multiple caustic formation. The   
fact that photon conservation does not forbid this has been explained, and an 
explicit, very detailed example of the basic underlying effect is presented 
in \cite{bi:shrink2}.  Paper II \cite{bi:shrink3} shows explicitly how 
`shrinking' (the increase in the area distance) occurs for 
compensated spherical lenses. These conclusions are supported by other 
studies, e.g. \cite{bi:dyer88,bi:HW97}, showing that most sources are 
demagnified rather than amplified when lensing occurs and caustics are 
taken into account. 

The increase of total wavefront area at high redshift ($z > 3$) 
is, however, strongly dependent on the model of the matter distribution used. 
Spatial averaging of nonlinearities, or the use of smoothed, linear 
matter distributions, may have a relatively mild effect on weak lensing 
\cite{bi:seljak}, but is known to strongly affect caustic formation and 
wavefront areas \cite{bi:HW97}.  
The critical issue underlying the effect we point out here, is that one 
is not allowed to smooth the matter distribution before calculating the null 
geodesics, because in the limit this excludes caustics. This explains the 
difference in  results between the weak lensing and Swiss-Cheese or 
semi-Swiss-Cheese \cite{bi:HW97} calculations. Our results
are hence in accord with those found in the Swiss-Cheese type models
\footnote{The 
tendency to see these models as unrealistic because of the exact matching 
conditions required in these models, is mistaken, in our view, because if 
the background model is correctly chosen, conditions of this kind {\it must}
be satisfied; see the discussion on compensation of lens overdensities 
in section 2 of this paper.}.

Caustics are expected to alter significantly observations of the CMB 
on small angular scales: principally they induce a non-Gaussian 
signature in the temperature anisotropies at the scale  
corresponding to the peak in the caustic distribution function . 
They will induce stronger changes to the angular
correlation function and hence the $C_{\ell}$ of the anisotropies than 
does weak lensing, simply because they involve non-perturbative mixing 
effects. It is just conceivable they could affect the spectrum at 
$\ell < 200$ due to multiple scatterings. If their effects do reach 
this far, caustics will alter the primary and 
secondary Doppler peaks, thereby contaminating parameter estimation 
programs \cite{bi:ref1}. 
    
In a previous draft of this paper, we suggested that 
the Dyer Roeder distance  might be usable on larger angular scales than 
generally supposed. This was criticized \cite{bi:seljak} on the grounds of 
photon conservation and neglect of shear. Here we propose an alternative 
interpretation of the Dyer Roeder distance - namely that it can be used to 
approximate the average area distance, including the shrinking due to 
caustics, {\em after all-sky}  averaging has been performed (thus giving 
it a validity in the opposite  regime to its normal implementation). This 
is based on the assumption  that caustics will be distributed in a 
statistically isotropic manner consistent with the symmetries of the matter 
correlation function. The Dyer Roeder  $\alpha$ parameter thereby  becomes 
related to the probability distribution of caustics and hence 
again back to the degree of inhomogeneity in the universe. 

In any case the main conclusion of the paper is that one should not 
assume the area-averaging result holds on large angular scales; 
rather the way angles relate to areas, and the consequent 
effect on observations, 
should be explicitly calculated for specific matter distributions. 
This paper and its companions show conclusively that the effect can occur.
How significant it is depends on the detailed matter distribution; it
will probably be small in most practical applications, but there might be 
circumstances where it is interesting.

\subsection*{ Acknowledgments}
The authors would like to thank Igor Barashenkov and Marco Bruni
for help at critical stages of writing. They would also like to thank Jurgen 
Ehlers, Uro$\breve{s}$ Seljak, Malcolm MacCallum, and Nazeem Mustapha  
for comments on earlier drafts, and the Caltech and Cambridge (UK)
lensing groups for comments
that led us to more realistic estimates of the effect than suggested in 
earlier drafts. BAB would like to thank Jeffrey Cloete for 
enlightening discussions over the years. We thank the FRD (South Africa) for 
financial support, and Mauro Carfora for drawing most of the diagrams.

\end{document}